\documentclass[journal]{vgtc}                     

\onlineid{1185}

\vgtccategory{Research}

\vgtcpapertype{please specify}

\title{\textit{Does This Have a Particular Meaning?} \\  Interactive Pattern Explanation for Network Visualizations}

\author{%
  \authororcid{Xinhuan Shu}{0000-0002-9736-4454},
  \authororcid{Alexis Pister}{0000-0002-2817-020X},
  \authororcid{Junxiu Tang}{0000-0003-3594-926X},
  \authororcid{Fanny Chevalier}{0000-0002-5585-7971},
  and \authororcid{Benjamin Bach}{0000-0002-9201-7744}
}

\authorfooter{
  \item
  	Xinhuan Shu is with Newcastle University.
  	E-mail: xinhuan.shu@ncl.ac.uk. The work was done when she was a postdoc at the University of Edinburgh.
  \item
  	Alexis Pister is with the University of Edinburgh.
  	E-mail: apister@ed.ac.uk.
  \item
  	Junxiu Tang is with Zhejiang University.
  	E-mail: tangjunxiu@zju.edu.cn.
  \item
  	Fanny Chevalier is with the University of Toronto.
  	E-mail: fanny@dgp.toronto.edu.
  \item Benjamin Bach is with Inria and the University of Edinburgh.
  	E-mail: benjamin.bach@inria.fr
}

\abstract{%
  This paper presents an interactive technique to explain visual patterns in network visualizations to analysts who 
  do not understand
  these visualizations and who are learning to read them.
  Learning a visualization requires mastering its visual grammar and decoding information presented through visual marks, graphical encodings, and spatial configurations. 
  To help people learn network visualization designs and extract meaningful information, we introduce the concept of \textit{interactive pattern explanation} that allows viewers to select an arbitrary area in a visualization, then automatically mines the underlying data patterns, and explains both visual and data patterns present in the viewer's selection. 
In a qualitative and a quantitative user study with a total of 32 participants, we compare interactive pattern explanations to textual-only and visual-only (cheatsheets) explanations. Our results show that interactive explanations increase learning of 
  \textit{i)} unfamiliar visualizations, 
  \textit{ii)} patterns in network science, and 
  \textit{iii)} the respective network terminology.
}


\teaser{
  \centering
  \includegraphics[width=\linewidth, alt={teaser.}]{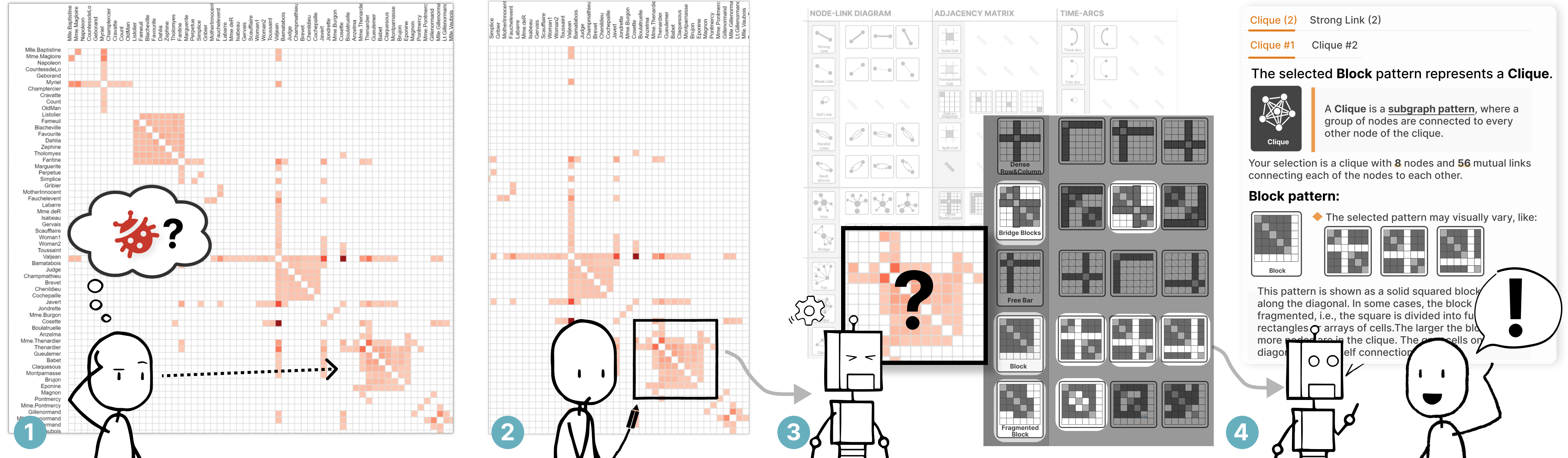}
  \caption{
  Pattern Explainer helps analysts unfamiliar with network visualizations learn about visual patterns in the representation of their data. \protect\circled{1} Looking at the visualization, a user spots a visual pattern of interest, e.g. a ``bug''-looking pattern in the matrix. To inquire about whether this pattern is meaningful, the user \protect\circled{2} selects the area. \protect\circled{3} Pattern Explainer then automatically mines the selection, against a dictionary of network motifs, and \protect\circled{4} provides the user with explanations of what underlying network patterns the visual pattern reveals.
  }
  \label{fig:teaser}
}

\graphicspath{{figs/}{figures/}{pictures/}{images/}{./}} 

\usepackage{tabu}                      
\usepackage{booktabs}                  
\usepackage{lipsum}                    
\usepackage{mwe}                       

\usepackage{mathptmx}                  

\usepackage{xspace}
\usepackage{multirow}
\usepackage{graphicx}
\usepackage[normalem]{ulem}

\begin{document}

\definecolor{ccolor}{RGB}{104, 175, 190}
\newcommand{\circled}[1]{\tikz[baseline=(char.base)]{\node[shape=circle,fill=ccolor,text=white,inner sep=0.6pt,outer sep=0pt, minimum size=0.5ex] (char) {#1}}}


\firstsection{Introduction}

\maketitle
A plethora of visualization designs have emerged for network analysis. Still, most of them are likely to be unfamiliar to the majority of analysts~\cite{zoss2018network,alkadi2022understanding}. 
Beyond the well-known node-link diagrams, such designs include adjacency matrices~\cite{henry2007matlink}, time-arcs~\cite{dang2016timearcs}, arc-diagrams, edge-bundling~\cite{holten2009force}, space-time cubes~\cite{bach2015networkcube}, and many more.
Reading these visualization designs requires learning the rules of their visual mappings and confidently decoding the marks' visual variables.
Ideally, the distance between the visualization as an external representation ~\cite{ware2019information} and a person's internal mental map of the concepts in the application domain should be narrow, allowing one to quickly read information from the visualization, search and inquire about specific information, understand new information, and perform more complex reasoning tasks such as insight discovery and decision making~\cite{boy2014principled}.

The matter is, analysts may face challenges in understanding unfamiliar visualization techniques, as visual encodings lead to abstract \textit{visual patterns} such as ``bugs'' in matrices (\autoref{fig:teaser}) or leaf shapes in time-arcs (\autoref{fig:visualizations}c).
Visual patterns are visually salient structures in a visualization that somehow catch an analyst's attention, while
interpreting these patterns can baffle those unfamiliar with the visualization technique. 
This obstacle was widely observed, for instance, when deploying the Vistorian network visualization tool~\cite{bach2015networkcube}, which provides multiple visualization alternatives, such as node-link diagrams, adjacency matrices, and time-arcs. 
Analysts from various backgrounds reported confusion and misinterpretation in perceiving visual patterns~\cite{alkadi2022understanding}. 
However, the different visualizations and their visual patterns ought to represent data characteristics and deliver insights, and aim to match each other, according to the fundamental principle of Representation Invariance for visualization~\cite{kindlmann2014algebraic,drucker2020visualization,pinker1990theory,andrienko2022seeking}.
Being able to quickly spot and interpret visual patterns is hence essential to using a visualization.

To support learning visualizations, we propose \textit{interactive pattern explanation}, explaining visual patterns and their corresponding topological network motifs on-demand in a user-defined part of the visualization.
As shown in \autoref{fig:teaser}, when exploring an unfamiliar network visualization, \circled{1} a user spots a visual pattern that they do not know how to interpret. 
The user \circled{2} selects the area of visual interest in that visualization.
Our approach automatically \circled{3} mines and retrieves all the underlying network motifs,
and \circled{4} presents visual-textual explanations.
Our explanations are similar to visualization cheat sheets~\cite{wang2020cheat}, using concise generic visuals and textual explanations characterizing these visual patterns. 
However, cheat sheets still require analysts' cognitive efforts to match canonical examples on the sheets with specific cases in the actual visualization.
Our premise is that \textit{visualizations are best learned experientially, i.e. by example and on-the-fly through an analyst exploring and analyzing their own data, rather than learning up-front and from generic descriptions in separate instructional materials.} 

In this paper, we focus on three representative network visualizations (node-link diagrams, adjacency matrices, and time-arcs) as a first step to study interactive pattern explanations on generic network patterns. 
We designed and implemented a proof-of-concept tool, Pattern Explainer, which can explain a repertoire of 34 visual patterns, across 3 visualization techniques, matching 11 common topological network motifs such as clusters, cliques, or hubs and 2 temporal network motifs (\autoref{fig:patterns}). Our 34 patterns were informed by a scoping study with four participants (\autoref{sec:study1}).
We compared Pattern Explainer to static cheat sheets in two qualitative/quantitative studies (\autoref{sec:evaluation}).
Our findings demonstrate that using the Explainer interface yields higher numbers of patterns people can correctly identify.  
Qualitative feedback suggests that participants appreciate the Explainer being intuitive and that integration within the visualizations provides in-context and on-the-fly explanations, recommending related instances for exploration, and validating peoples' understanding.
Supplementary materials contain a demonstration video, pattern detection methods, and study materials.
\section{Background}

We summarize concepts around patterns in visualization, patterns in networks (in particular motifs), and visualization teaching approaches.

\subsection{Network Exploration and Network Literacy}

Empirical studies demonstrate a lack of knowledge about networks among the general population~\cite{borner2016investigating,zoss2018network}. For example, Alkadi et al.~\cite{alkadi2022understanding} identified eight general barriers to visual network exploration, such as interpreting visual patterns and developing trust in a given visualization. Another study, showing images of node-link diagrams to visitors of a science museum, found that most visitors struggled to explain how to interpret these images~\cite{borner2016investigating}. Such studies support recent calls for \textit{network literacy}~\cite{sayama2016essential, cramer2015netsci} and \textit{network visualization literacy}~\cite{zoss2018network} which include notions of `networks help reveal patterns', `networks can be visualized in many different ways'~\cite{cramer2015netsci,anderson2001taxonomy} as well as \textit{topological literacy}. 

While networks can contain many different information and data patterns, for simplicity, we focus on the notion \textit{topological} structures that characterize subgraphs or motifs. Subgraphs mean arbitrary subsets of nodes and links, whereas motifs are subgraphs with a meaning which can recur in a network. Motifs can consequently be analyzed, compared, and mined, to support different network exploration tasks~\cite{lee2006task, bach2013graphdiaries, ahn2013task}. Common network motifs include star motifs, triangles, clusters, cliques, circles, bridge nodes, fans, and many more~\cite{newman2018networks}. The number of such motifs is theoretically infinite, and different sets of motifs are described and used across different domains (e.g.,~\cite{wassermanSocialNetworkAnalysis1994, carringtonModelsMethodsSocial2009, przuljBiologicalNetworkComparison2007}). 
Numerous mechanisms are proposed to query and identify the respective topological structure from networks (e.g.,~\cite{schreiber2005mavisto,lekschas2017hipiler, pister_combinet_2023}). 

Unfortunately, only a few network visualizations come with precise explanations of the \textit{visual patterns} that topological motifs look like, when being visualized. 
Behrisch et al.~\cite{behrisch2016matrix} described five patterns for adjacency matrices (Block, Off-diagonal Block, Line/Star, Band) and two anti-patterns (Noise and Bandwidth).
Follow-up work~\cite{behrisch2016magnostics} applies image feature recognition to quantify the visual quality and saliency of these patterns. Likewise, Bach et al.~\cite{bach2016towards} describe visual patterns for graph motifs (clusters, cliques, hubs, stars, etc.) 
for Confluence Graphs. To promote network literacy, we contribute an initial dictionary of visual patterns and associated topological motifs for three network visualizations, 
which informs our interactive pattern explanation.

\begin{figure*}
    \centering
    \includegraphics[width=\textwidth]{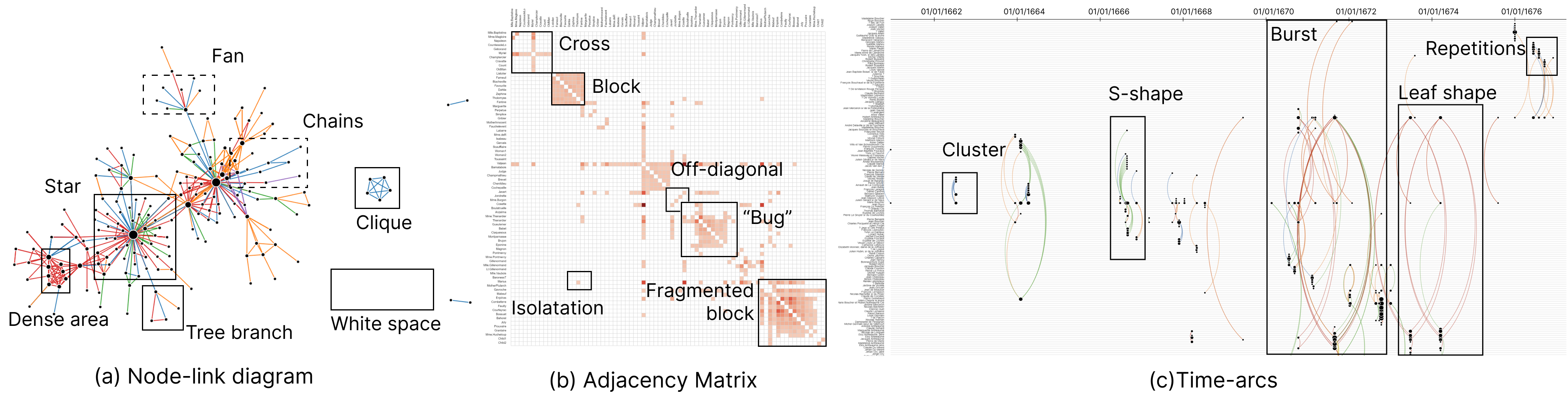}
    \caption{Network visualizations considered in this work and their particular designs.
    The solid rectangles show some portions participants annotated, while the text around was added by the authors for clarity in the paper, not visible to the participants. 
    The dashed rectangles show some visual patterns the authors annotated. 
    The designs in these visualizations are optimized by the authors of this paper for readability. 
    (a) and (c) use the \textit{Marie Boucher Trade} network dataset~\cite{marieBoucher}, and (b) uses the \textit{ `Les Mis\'erables' Co-occurrence} network dataset~\cite{netRep}.
    }
    \label{fig:visualizations}
    \vspace{-0.3cm}
\end{figure*}

\subsection{Visual Patterns and Visualization Literacy}
\label{sec:literacy}

Visual patterns in data visualizations are often described in the context of \textit{``trends and outliers''}~\cite{munzner2014visualization, ware2019information}, or as visually salient areas that catch a person's attention and which result from the groupings described by the Gestalt Laws such as similarity, proximity, common fate~\cite{koffka1922perception}. Likewise, patterns can be described as the \textit{``co-location of particular visual features''}~\cite{andrienko2022seeking} or a \textit{``meaningful whole''}~\cite{collins2018guidance}, that is, multiple data elements with a special spatial arrangement. 

Andrienko et al.~\cite{andrienko2022seeking} described a comprehensive framework for visual patterns for Visual Analytics. In that framework, \textit{``a pattern consists of relationships between multiple elements of at least two data components''} which are \textit{``represented as a single object as, for example, a cluster, trend or correlation.''} The framework further describes different types of patterns, the notion of sub-patterns as well as overlay between patterns. In viewing a visualization, a viewer \textit{``detects salient visual patterns, [and] translates them into conceptual information structures''}~\cite{andrienko2022seeking}, i.e., a visualization can be thought of as being subdivided into \textit{visual chunks}~\cite{lowe1988reading} which are perceived, aggregated, and abstracted into higher-level (information) chunks and more complex patterns. Eventually, these patterns are associated with meaning and included into one's mental models~\cite{collins2018guidance} for higher-level cognitive processes, such as deriving insights and making decisions. This way of reading a visualization is commonly referred to as \textit{bottom-up} and implies exploration, hypothesis generation, and even serendipity.
Opposed to that is a \textit{top-down} approach which emphasizes leveraging a viewer's knowledge to selectively perceive relevant visual patterns~\cite{freedman2002toward, pinker1990theory}. 

Visual patterns are thus a crucial aspect of reading visualization. 
Speaking of visualization literacy as \textit{``the ability to make meaning from and interpret patterns, trends, and correlations in visual representations of data''}~\cite{borner2019data} includes the notion of visual patterns, owing strong ties to related concepts of graphicacy~\cite{wainer1980test} and visual literacy~\cite{bristor1994linking, chevalier2018observations}.
Specifically, Boy et al.~\cite{boy2014principled} described visualization literacy as \textit{``the ability to confidently use a given data visualization to translate questions specified in the \textbf{data domain} into visual queries in the \textbf{visual domain}, as well as interpreting \textbf{visual patterns} in the visual domain as properties in the data domain.''} (emphasis added). This translation is exactly what Pattern Explainer aims to help with.

\subsection{Interactive Learning of Data Visualizations}
Learning novel visualization techniques is a core challenge in visualization education~\cite{bach2023challenges}, despite a variety of resources available~\cite{liu2023visualization}.
For example, good resources incorporate concrete examples in textual descriptions~\cite{ stoiber2022abstract}, activities and exercises~\cite{tanahashi2016study, alper2017visualization}, games~\cite{huynh2020designing}, animations between alternative visualization techniques~\cite{ruchikachorn2015learning, henry2007nodetrix}, as well as interactive tutorials for parallel coordinates~\cite{kwon2016comparative, firat2022p, stoiber2021design} and treemaps~\cite{firat2020treemap}. 
However, these approaches consist of separate instructional materials that are limited in providing in-situ help when novices explore their own data and have difficulty in interpretation. 
Similar issues exist in other user interface research, and have motivated tools such as GestureBar~\cite{bragdon2009gesturebar} which discloses relevant gesture details only as needed while users work with tools.
It is yet underexplored for visualization learning.

While these materials provide general knowledge about visualization techniques, one example has looked more deeply into how to interpret patterns embodied by the visualization. Namely, visualization cheat sheets~\cite{wang2020cheat} explain visualization techniques, their visual encodings, and their visual patterns through well-defined, handcrafted, concise visual and textual descriptions. Cheat sheets are meant to be consulted both \textit{up-front} or \textit{during} an exploration task. Still, much like the above cases, cheat sheets require a cognitive translation from the canonical pattern examples to the analyst's visual patterns created by their data at hand.

We study interactive pattern explanations that inquire and explain visual patterns \textit{on-demand}, as these patterns emerge from the analyst's data. Our approach is different from visual query interfaces~\cite{wattenberg2001sketching, fan2020sketch, bhowmick2020aurora} which require a user to know more or less precisely what they are looking for. 
Pattern explanation starts from an existing instance of a suspected pattern, and provides just-in-time interactive learning.
This approach is more akin to in-situ help for user interfaces~\cite{chundury2023contextual} and guidance and recommender engines for data analysis~\cite{li2023networknarratives}.
\section{Scope and Challenges}
\label{sec:challenges}

This section engages in a deeper discussion of the challenges (Ch1-Ch4) in trying to define visual patterns. 

\subsection*{Ch1: How to define the concept of visual patterns}
\label{sec:definition}

The exact meaning of the notion of visual patterns in visualizations is hard to define in a generic sense. Its meaning can be varied in different contexts, and what exactly makes an individual pattern will highly depend on a) the visualization design, the data, visual contrast, and the presence and shape of other patterns, as well as b) an observer's experience, knowledge, and task. For the purpose of this paper, we define a visual pattern as a \textit{spatially contained, salient configuration of visual marks in a visualization that attracts an observer's attention}. This definition tells us \textit{what} an observer might find interesting (a region in the visualization), independently from \textit{why} they find it interesting (beyond the scope of this paper).
Examples of visual patterns include a set of close points in a node-link diagram (\autoref{fig:visualizations}a), a block of colored squares or an outlier isolation in a matrix visualization (\autoref{fig:visualizations}b).

\subsection*{Ch2: What visual patterns appear in network visualizations}

For this research, we focus on three representative network visualization techniques whose designs we have carefully developed to clearly show visual patterns (\autoref{fig:visualizations}).

\begin{itemize}[noitemsep, leftmargin=*, partopsep=0pt,parsep=\parskip,topsep=1pt]
\item \textbf{Node-link diagrams} (\autoref{fig:visualizations}a): Common visual patterns in node-link diagrams include stars, fans, chains, dense areas. Link color and thickness in our designs represent link types and link weight, respectively. The saliency of many of these visual patterns depends on the node positions in the respective graph layout. In our study, we use the default WebCoLa~\cite{cola} layout to minimize the distance between connected nodes and ensure nodes do not overlap and are spaced apart appropriately. 
    
\item \textbf{Adjacency matrices} (\autoref{fig:visualizations}b) are likely to be less familiar to many people~\cite{ghoniem2004comparison}. Matrices can create very peculiar visual patterns such as crosses, blocks, fragmented blocks, outliers, lines, or all sorts of insect-like looking patterns along the diagonal (\autoref{fig:teaser}). Like in node-link diagrams, these visual patterns depend on the algorithm dictating the layout of rows and columns, called the matrix seriation method~\cite{behrisch2016magnostics,behrisch2016matrix}. In our design, we use the Barycenter heuristic implemented in Reorder.js~\cite{fekete2015reorder} to order rows and columns. This algorithm optimizes the ordering of the nodes so that every node is the closest possible to its topological neighbors in the matrix ordering. Cell shading (darkness) maps to link weight.

\item \textbf{Time-arcs} (\autoref{fig:visualizations}c) visualize temporal networks~\cite{dang2016timearcs} in a Cartesian coordinate system with a time axis (x-axis, left-to-right) and a list of nodes (y-axis). A connection between two nodes in time is shown as an arc connecting two circles placed at the intersection of the node and the point in time when the connection happens. Arcs encode link direction, counter-clockwise from the start to the end node. Visual patterns in time-arcs include overlapping arcs, recurring arcs, rows with lots of dots, sets of arcs resembling a S or leaf shape, clustered arcs, etc. Similar to matrices, the saliency and specific appearance of visual patterns in time-arcs depend on the node ordering on the y-axis. This ordering can make arcs spanning the whole height of the visualization and flip the direction of an arc depending on the link direction. To order nodes along the vertical dimension, we used the same Barycenter seriation method in the matrix visualization.
\end{itemize}

\subsection*{Ch3: How to formalize network patterns}
\label{sec:Ch3}

Our second definition is that of a \textit{network pattern} to describe any \textit{topological structure in a data set with some meaning}. The meanings of these patterns can be described through network motifs: clusters, cliques, bi-cliques, bridge-nodes, pathways, etc. For simplicity, in this paper, we focus only on mappings of visual patterns to network motifs, excluding more general information in networks such as density, homogeneity, and symmetry.

\subsection*{Ch4: Mapping visual to network patterns}
\label{sec:Ch4}

Given visual and network patterns, a visualization creates \textit{a possible mapping from a visual pattern to a data pattern and vice-versa, determined by the rules of the particular visual mapping.} This definition says that a visual pattern implies, but does not equate the existence of a corresponding data pattern (and vice versa). There are different possibilities for such mappings.

\begin{itemize}[noitemsep, leftmargin=*, partopsep=0pt,parsep=\parskip,topsep=1pt]
    \item The ideal situation is a \textbf{one-to-one mapping}  
    where a visual pattern (VP) is mapped to a network pattern (NP) and vice versa: $VP \leftrightarrow NP$. That would mean that the presence of a visual pattern necessarily indicates the presence of a network pattern (and vice versa). Properties of the visual patterns (location, size, structural characteristics) could then be used as a proxy to interpret the data pattern.
    
    \item A visual pattern can also be an \textbf{artifact} of the layout algorithm or visual encoding and not map to any meaningful pattern in the data: $VP \dashv NP$. Examples of artifacts in matrices include band patterns or the ``bugs'' that can occur in specific seriation methods but do not relate to any particular network motif~\cite{behrisch2016matrix}. Likewise, in \autoref{fig:visualizations}c we can see repetitive dot patterns that again result from node seriation, but do not translate to a meaningful topological feature.    

    \item Likewise, a visualization can \textbf{obscure} a pattern present in the data, i.e., not resulting in any easily perceivable visual pattern $VP \vdash NP$. Examples of such obscurations include cliques or clusters in matrices that happen to not appear as block patterns but instead the individual cells are spread over the matrix (\autoref{fig:overview}~\circled{5}-Clique \#1).
    
    \item A special case of artifacts are \textbf{confusers} where one visual pattern implies multiple possible patterns (explanations) in the data $VP \leftrightarrow \{NP_1,NP_2, ...\}$, including wrong ones. Examples of such cases include links overlapping nodes in a node-link diagram but not connecting to these nodes~\cite{wong2007supporting}; or, dense areas in node-link diagrams that originate from incidentally overlapping nodes and links but do not form a topological cluster. 

    \item \textbf{Hallucinators} are seemingly different visual patterns that map to the same network pattern $\{VP_1, VP_2, ...\} \leftrightarrow NP$. 
    Examples include any visual pattern impacted by layout, especially orderings in time-arcs where some arcs are short, others are long; or a block in a matrix becoming a split block with a mere reordering of the rows. 
\end{itemize}

Kindelman and Scheidegger~\cite{kindlmann2014algebraic} provide the framework of Algebraic Visualization Design to think about these different mappings and their implications. In that framework, \textbf{structural variation} between two data sets 
are denoted by the abstract distance $\alpha$ whereas the \textbf{visual variation} $\omega$ is the visual distance between two visual patterns. 
In other words, two clusters show structural variation if they are not graph-isomorph~\cite{whitney1992congruent}. Structural variation is supposed to be reflected in the visual patterns and allows for the comparison and characterization of network patterns. Visual variation between visual patterns includes purely visual differences in a visualization ($\omega>1$) while showing the exact same pattern instance ($\alpha=0$). 

A good visualization design equates any changes in $\alpha$ with a change in $\omega$. In fact, our above terminology of Hallucinators and Confusers is taken from Kindelman and Scheidegger~\cite{kindlmann2014algebraic}: for any $\alpha >0$ not resulting in a visual difference in the visualization $\omega = 0$, the visual patterns are Confusers. Likewise, any difference in $\omega$ without a difference in the data $\alpha$ is called a Hallucinator.
Another way of thinking about variation is to use Plato's framework of \textbf{ideas} (of a pattern) and their \textbf{instance(s)}. The ideas of patterns are those meanings and terms we attribute a class of instances of these patterns: `block', `cross', or `group of nodes' as examples for visual patterns and `clique', `cluster', `bridge node' for network patterns (motifs). Instances are the particular occurrences of such an idea, each of which may vary significantly from each other. 

Pattern Explainer covers all the above situations, explaining \textit{one-to-one mappings}, \textit{artifacts} (displaying a message that the selected visual pattern is likely to be an artifact, since no corresponding network pattern could be found), \textit{confusers} (showing all underlying network patterns), \textit{obscurations} (highlight all network patterns present in the visualization), and \textit{hallucinators} (showing related visual patterns).
We refer back to these challenges in the following sections, justifying our design choices and contextualizing findings from our studies. 
\section{Scoping Study: Understanding People Reading Network Visualizations}
\label{sec:study1}

We conducted a first user study to help us scope our research and investigate
\textit{\textbf{Q1}: What visual patterns do people see in network visualizations?}, and \textit{\textbf{Q2}: Which of these visual patterns do require explanations}?

\subsection{Study Design}

We recruited 4 participants from a local university in the UK. 
Three participants (P1-P3) were Master students in Design and Computer Science, while one was a postdoctoral researcher in social science (P4). 
All participants had a basic understanding of networks but only knew about node-link diagrams. All participants were volunteers, unpaid. 
We created two visualization examples for each of the following network visualizations: node-link diagram, adjacent matrix, and time arcs, totaling six visualizations (\autoref{fig:visualizations}). Visualization designs were the same as detailed in \autoref{sec:challenges}.
The selection of these instances aimed to cover a variety of data and visual patterns. 
The study materials can be found in supplementary materials. 

Sessions were held individually in person and lasted around 45 min each. Participants first filled out a consent form and a demographic questionnaire, then, we asked them to read and annotate each of the six visualization examples, one at a time. For each visualization, we provided a textual explanation of visual encodings, which were available to participants throughout the entire study. 
To alleviate the effects of specific terminology, we described all of the networks as social networks with `people' referring to nodes and `connection' referring to links.
For each visualization, we asked participants to tell us:
\textit{1)} \textit{everything they've learned or feel is meaningful in this network}, 
 and \textit{2)} \textit{anything that confused them and required more explanations}, 
while annotating the corresponding areas.  
Once they were done annotating all visualizations, we interviewed each participant by asking what confused them most regarding each visualization technique and what explanations they would like. 
The study was run on an 11'' iPad Pro operated with an Apple Pencil, with video- and screen-recording on.

\subsection{Findings}

Two major observations referring to patterns participants could and could not understand emerged from our study. 

\textbf{Observations 1: Participants did not look at patterns other than clusters, highly connected nodes, and strong links (Q1).}
Generally, participants started with fundamental information, e.g., node distribution, link density, dominant node or link types, and dense areas.
The patterns they annotated were mostly strong (thick) links, highly connected nodes, and clusters, which they explained using their own words. 
For example, for highly connected nodes, P2 said, \textit{``This is a large node, and it connects to many nodes.''}, and P1 described clusters as: \textit{``This is a dense area. Nodes have lots of connections with each other''}.
Only P4 pointed out a bridge-node pattern in a node-link diagram, verbalizing \textit{``This node connects these two groups''}.

\textbf{Observation 2: Participants wanted clarification for different visual patterns (Q2).}
After longer examination, participants reported recurring salient visual patterns, such as radial shapes in node-link diagrams, dense areas in matrices, and leaf shapes in time arcs (\autoref{fig:visualizations}). 
However, they were confused about the meaning of many of the patterns and their particular differences even though they share some similarities: \textit{``Why is the structure in the dense area different from those radial structures''}(P1) (\autoref{fig:visualizations}a).
Circling a bug-like area (cluster) in a matrix (\autoref{fig:visualizations}b), P2 asked \textit{``It looks like a bug. Does it have a particular meaning? Or just happen to [have no meaning]?''}
P3 could understand that a leaf shape in time arcs (\autoref{fig:visualizations}c) meant that the two nodes had connections mutually pointing at each other, but they were confused about the length of the respective arcs \textit{``Why is this leaf[-shape] larger than that [other one]? Does that mean [the connection lasts] a longer time?''}
Additional confusion occurred due to complex visuals, which were likely a mixture of multiple patterns (\autoref{fig:visualizations}c-Burst). Participants also found it hard to discriminate specific visual patterns because of visual clutter and noise obscuring particular patterns. 
Besides, P2 asked about white space in the node-link diagram, \textit{``Does the distance [white space] mean they are less relevant?''}.

There was consensus that node-link diagrams were the easiest to understand. Matrices were less easy and caused confusion because of their symmetry and the resulting patterns. Time-arcs were found most difficult as they required understanding networks changing over time. 
\section{Pattern Explainer}
\label{sec:system}

\begin{figure*}[htb]
  \includegraphics[width=\textwidth]{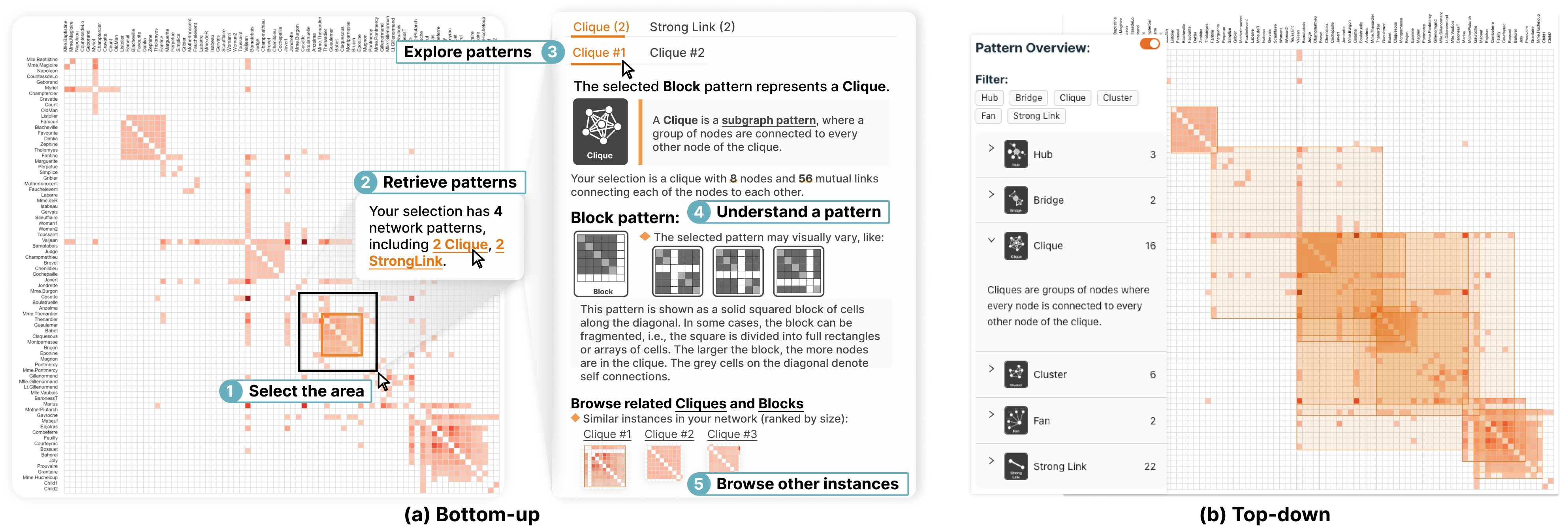}
  \caption{
  The Pattern Explainer idea. (a) In the bottom-up explanation, a user can \protect\circled{1} select an arbitrary region of interest in a network visualization. Our system \protect\circled{2} retrieves all the underlying network patterns in the user selection, backed by a pattern repository and a set of heuristics, and pops up an overview. After a user \protect\circled{3} selects a pattern for exploration, the pop-up \protect\circled{4} provides visual-textual explanations of the network and visual patterns, and \protect\circled{5} lists other instances for browsing. 
  (b) In the top-down explanation, the user can browse all the found instances according to pattern types.
  }
  \vspace{-0.3cm}
  \label{fig:overview}
\end{figure*}

Pattern Explainer is a proof-of-concept interface we developed to explore and evaluate interactive pattern explanation. The interface (\autoref{fig:overview}) mainly contains a network visualization and can in theory be wrapped around any network visualization application. 
Interactive pattern explanation with our interface works as follows (blue-circled labels refer to parts of \autoref{fig:overview}a): 

\begin{itemize}[noitemsep, leftmargin=*, partopsep=0pt,parsep=\parskip,topsep=1pt]
    
    \item 
    As part of an analysis task, a user explores the visualization and spots a visual pattern which they do not know how to interpret.
    
    \item 
    The user then \circled{1} selects the respective part of the visualization through a rectangle or lasso selection. This selection essentially defines a subgraph containing the nodes and/or links represented by the visual marks in the selection.
    
    \item 
    Then, Pattern Explainer \circled{2} looks for all possible network patterns in this subgraph, using a pattern repository (\autoref{sec:patterns}) and a set of heuristics when a motif qualifies as such (\autoref{sec:patterndetection}).  It may happen that the visual pattern selected by the user refers to multiple network patterns (confusers). 

    \item The \textbf{selector pop-up} shows up with the list of network patterns found, indicating, e.g., \textit{``Your selection has 4 network patterns, including 2 cliques and 2 strong links.''} Likely, if no patterns are found (artifacts), a message informs the user, explaining that the selected visual pattern is most likely an artifact. 

    \item Clicking on either of the patterns in the pop-up brings up the \textbf{explainer pop-up} (\autoref{sec:popup}).
    The explainer pop-up \circled{3} explains the chosen network pattern as well as \circled{4} the corresponding visual pattern through corresponding visual and textual descriptions.

    \item 
    After reading in the explainer pop-up, a user can \circled{5} continue either by 
    \textit{i)} exploring other instances of the same network pattern in the network, 
    \textit{ii)} choosing another network pattern from the selector pop-up, or 
    \textit{iii)} selecting another visual patterns in the visualization.

\end{itemize}

We call this process \textit{bottom-up explanation}, because a user starts from a single particular visual pattern in the visualization. 
The interactions \textit{offer informative feedback} and \textit{design dialogs} with a sequence of actions, following the user interface design rules~\cite{shneiderman2010designing}.

Pattern Explainer can also support \textit{top-down explanation} in which it first retrieves \textit{all} network patterns in the entire visualization, and highlights them alongside a summary of network patterns and their quantities (\autoref{fig:overview}b). In the top-down explanation, a user can immediately start browsing patterns, complementary to bottom-up explanation. 
They use the same explainer pop-up to keep the consistency~\cite{shneiderman2010designing}.
A toggle button in the interface switches between both explanation modes.

The remainder of this section details the design and rationales for the explainer pop-up, the pattern repository, and the pattern detection.

\begin{figure*}[htb]
    \centering    
    \includegraphics[width=\textwidth]{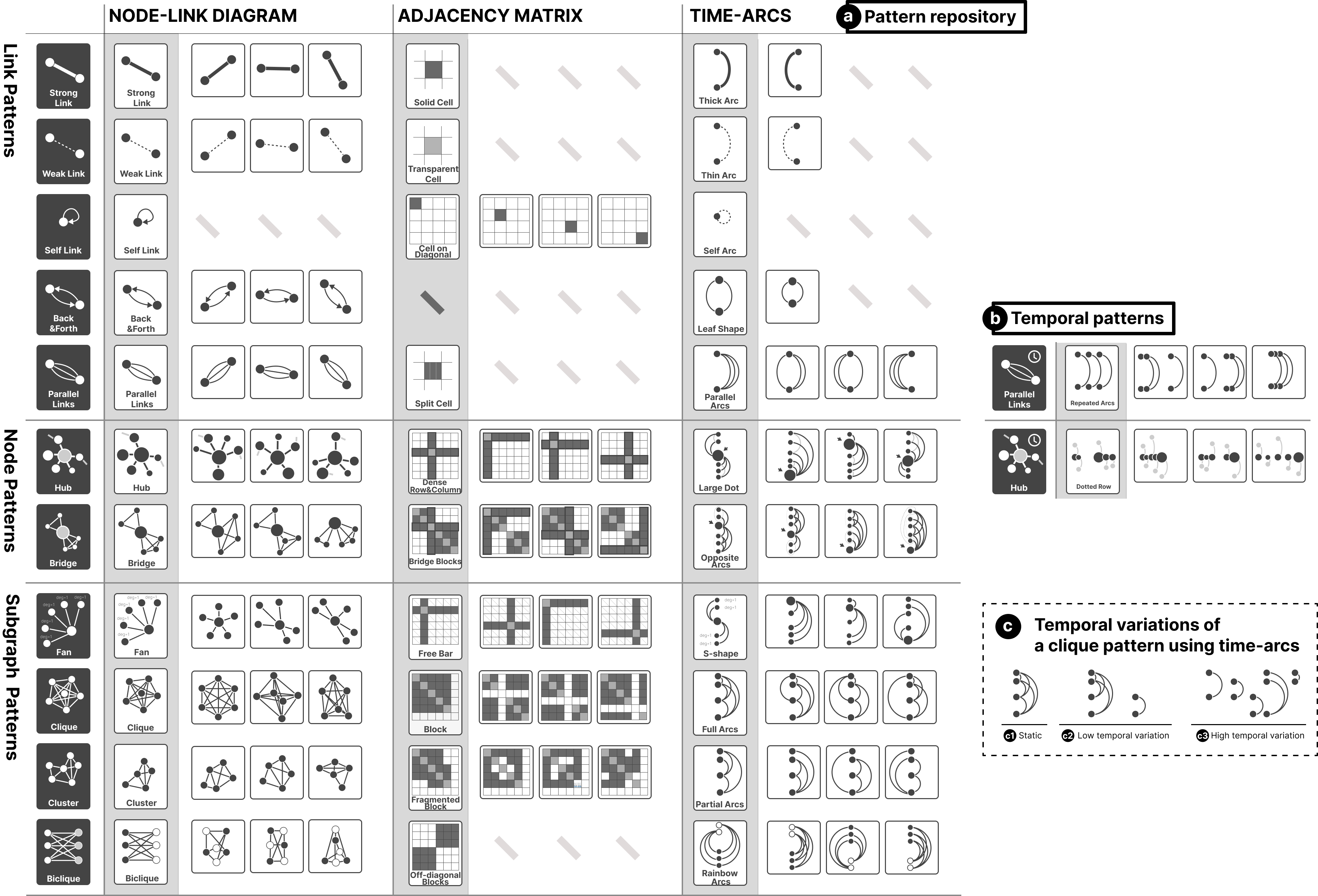}
    \caption{An overview of the pattern repository (a, b). Network patterns (black background) are organized vertically, while corresponding visual patterns (white background) are listed horizontally in each of the three visualizations. Icons for visual patterns include one lead icon (gray background) and several smaller versions of visual variations. (c) illustrates temporal variations of a clique pattern in time-arcs that were excluded in the repository. 
    } 
    \label{fig:patterns}
    \vspace{-0.3cm}
\end{figure*}

\subsection{Explainer Pop-up}
\label{sec:popup}

\begin{figure}[htb]
    \centering
    \includegraphics[width=\linewidth]{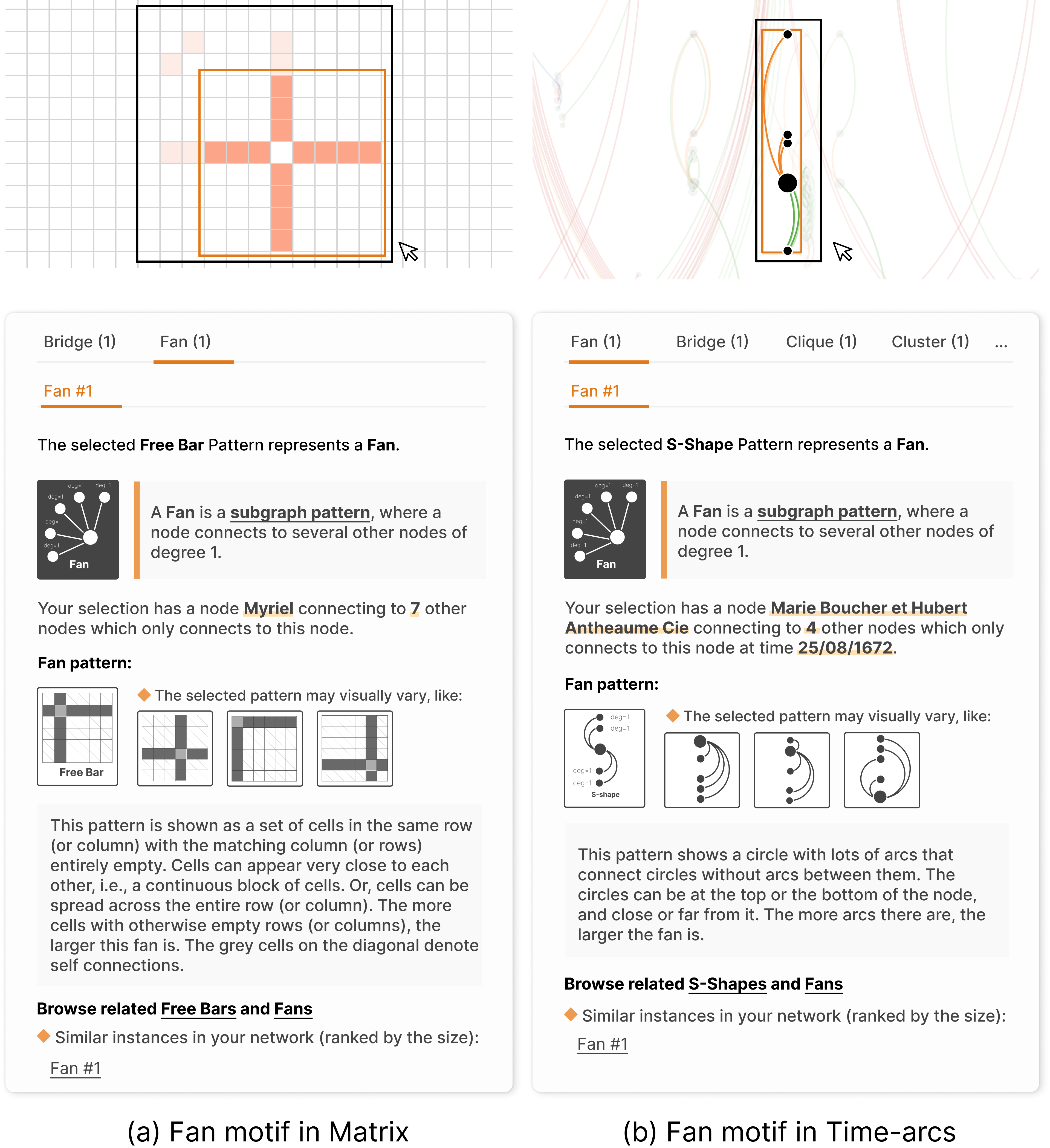}
    \caption{Explainer pop-up: Example of selections and generated pop-ups for the fan motif in (a) a matrix and (b) a time-arcs visualization. }
    \label{fig:pop-up}
    \vspace{-0.3cm}
\end{figure}

The explainer pop-up (\autoref{fig:overview}\circled{3}-\circled{5}, \autoref{fig:pop-up}) is shown when a user chooses a network pattern in the selector pop-up (bottom-up explanation) or selects a highlighted pattern while in top-down explanation. The explainer pop-up is inspired by visualization cheatsheets~\cite{wang2020cheat} and was iterated based on feedback from authors and participants in evaluation piloting. The pop-up consists of six components:

The \textbf{tab menu} on the top lists all network patterns identified in the user selection visual pattern (confusers). Like in the selector pop-up, a user can chose which pattern to explain.

Second, the explainer pop-up shows a \textbf{black icon illustrating the network pattern}.
These graphical icons represent network concepts using the most intuitive form of nodes and links, echoing findings from our scoping study (\autoref{sec:study1}).
Complementary textual explanations use abstract and concise language, e.g., \textit{``A Fan is a subgraph pattern, where a node connects to several other nodes of degree 1''}.

Third, a text explains \textbf{facts about the network pattern} as it appears in the network, e.g., the number of nodes and links, link density, and link weight ranking. These data facts aim to help understand structural variation in network patterns and to link the abstract idea of a pattern to the selected data (\hyperref[sec:Ch4]{Ch4}). These facts are inspired by more general network facts in  NetworkNarratives~\cite{li2023networknarratives}.

Fourth, the \textbf{visual pattern is explained} showing a visual icon on a white background that represents the idea and its visual pattern: each of the three visualizations has its own set of visual icons (see \autoref{fig:patterns}). Visual pattern icons on a white background distinguish them visually from the network patterns. As we did not find any established terms for these patterns, we initiated our own names. Participants in the evaluation indeed used these names to reference those visual patterns. Textual explanations below the icons explain the idea and characteristics of the visual patterns, as shown in \autoref{fig:pop-up}a for a fan pattern in a matrix visualization (Free Bar). Like visual pattern names, explanatory visuals and texts have been refined through discussion among all authors and integrated feedback from participants in a pilot study.

Fifth, \textbf{visual variations of the visual pattern} are shown as complementary array of hand-crafted icons to show possible visual variations of the visual pattern in this visualization (Ch4, Hallucinators, \autoref{fig:patterns}). The three variation icons have been designed to show different aspects of visual variation in the visual patterns while still mapping to exactly the same network pattern. 

Last, a list of previews to \textbf{related pattern instances} is shown, linking to other instances of the same network pattern in the same network. These pattern instances are retrieved up-front through the same mechanism as in top-down explanation. Hovering any pattern instances in the explainer pop-up highlights the respective instance in the visualization. This feature aims to help people further a user's understanding of structural and visual variations by example (\hyperref[sec:Ch4]{Ch4}).  

\subsection{Pattern Repository and Pattern Detection}
\label{sec:patterns}

Pattern Explainer currently hosts 11 network patterns and 34 visual patterns across node-link diagrams, matrices, and time-arcs (\autoref{fig:patterns}). 
To create this repository (\hyperref[sec:Ch3]{Ch3} \& \hyperref[sec:Ch4]{Ch4}), we adopted the following process.

\begin{enumerate}[noitemsep, leftmargin=*, partopsep=0pt,parsep=\parskip,topsep=1pt]
\item For each of our three visualizations, we chose an initial set of visual patterns that were salient and observed by participants in the scoping study and ourselves (\autoref{fig:visualizations}).
This resulted in about 4-10 visual patterns per visualization and three sets of visual patterns $V_1, V_2, V_3$. 

\item Then, we \textbf{associated these visual patterns to their corresponding network patterns}. This process resulted in three network pattern sets $N_1, N_2, N_3$, one for each visualization.
To simplify, we adopted a 1:1 mapping from each visual pattern in $V_i$ to one network pattern in $N_i$ with $|V_i| = |N_i|$. 
\item Across all three visualizations, this process defined our set of network patterns $N = N_1 \cap N_2 \cap N_3$. 
\item We \textbf{discussed the set of network patterns N}, and discarded several patterns whose visual patterns were hard to detect by people most of the time, such as the bipartite pattern.
We kept the special case of the biclique pattern, which often appears more salient, especially in matrices. 
We also agreed to remove most of the temporal patterns from the time-arc visualization (bursts, repetition, temporal clusters, etc.). These network patterns presented a very high visual variability and unclear heuristics for their detection (\autoref{fig:patterns}c).
\item In the last step, we checked for any network pattern in N that did not have a corresponding visual pattern in each visualization, i.e., $V_1, V_2, V_3$. This led to defining new visual patterns for some of the visualizations which were not part of our initial creation of $V_1, V_2, V_3$. In summary, we ended up with $|V_i|=|N|$ for all visualizations $i$.
\end{enumerate}

\subsection{Network Pattern Detection}
\label{sec:patterndetection}

To detect network patterns in user selection (bottom-up explanation) and the entire network (top-down explanation), we implemented a set of existing algorithms for network motif detection (e.g.,~\cite{dunneMotifSimplificationImproving2013, clique1, louvain}). 
We defined and refined a set of heuristics when a motif qualifies as a pattern to refine the detection. 
For example, a cluster or clique must have at least 5 nodes; a fan must have at least 4 nodes. A full description of the detection algorithms and heuristics we used, corresponding pattern explanations, and benchmark analysis are provided in the supplementary materials. Our detection methods can be applied to any network datasets and easily amended to fit additional network patterns.
\section{Evaluation}
\label{sec:evaluation}

We conducted two complementary user studies to collect qualitative and quantitative data about how participants use pattern explanation to learn network visualizations. We hypothesized interactive explanation increases the learning of visualizations and network patterns, compared to baseline techniques such as text-only explanations or static textual-visual explanations on visualization cheatsheets.

\subsection{Qualitative Study}
\label{sec:study2}

We adopted a within-subjects design to collect participants' qualitative feedback on three different techniques (i.e., textual explanations, cheat sheets, and Pattern Explainer) for interpreting network visualizations.
The study is deliberately designed in an open-ended way, as any study about learning inevitably alters its subject of study.
We cannot interrogate the same subject twice about their learning outcomes.

\subsubsection{Study Setup} 

\textbf{Participants}---We recruited participants by disseminating advertisements through emails and online social groups.
Participants were expected to know little to nothing about the tested network visualizations, and we asked them to self-report their knowledge
of these visualizations on a 5-point Likert scale (never seen (1) to expertise (5)). 
As a result, we selected 12 participants (P1-P12)
(6 males and 6 females, aged ranging from 20 to 29, mean: 24.8) 
with diverse backgrounds (law (1), sociology (1), design (1), environmental engineering(1), project manager (1), veterinary medicine (1), chemical engineering (1), mechanical engineering (1), computer science (2), and software engineer (2)).
Most participants reported they had not seen adjacency matrices and time arcs before (4 had seen but not used matrices, and 2 had seen but not used time-arcs). 
Node-link diagrams were more familiar: 10 participants had seen but not used, 1 had used, and 1 had expertise. 

\textbf{Data}---We use \textit{Marie Boucher Trade} network dataset~\cite{marieBoucher} (189 nodes, 488 links, 91 patterns identified in Pattern Explainer) for node-link diagrams and time-arcs, and the \textit{ `Les Mis\'erables' Co-occurrence} dataset~\cite{netRep} (77 nodes, 254 links, 43 patterns) for adjacency matrices.

\textbf{Conditions}---We designed three conditions with differential information, i.e., only textual explanations of visual encoding, cheat sheets, and Pattern Explainer.
The text-only condition serves as a baseline to examine if additional instructions about patterns are needed for interpretation. 
We created three A4 cheat sheets, one for each visualization. On each sheet, we added screenshots of the explainer pop-ups for each network pattern, removing data facts and related pattern instances, since cheat sheets only provide visual and textual explanations of patterns with canonical examples.
In the Pattern Explainer condition, we only provide the bottom-up mode (\autoref{fig:overview}a) to investigate the effects of our interactive in-situ explanations and pattern instances, compared to static and premeditated cheat sheets.

\textbf{Procedure}---We followed a structured interview process. 
Participants started with consent and a brief introduction.
Then, we asked each participant to interpret node-link diagrams, adjacency matrices, and time-arcs in three different conditions.
To alleviate the order effects, we counterbalanced the order of visualizations and conditions.
We deliberately set the text-only explanations to always be the first condition, as this condition would not truly exist if they had used cheat sheets or Pattern Explainer first.

For each visualization, participants were given a limited time based on pilot trials, 5 minutes for the text-only condition and 10 minutes for the other two to go through the explanations and functions.
They should read and interpret using the given intervention, and respond to our prompt: \textit{``Tell us what you learn about the network and anything you feel is interesting or important about the network, and refer to the visualization.''}
They were required to think aloud during the process. 

Once the 3 visualizations were completed, we interviewed participants with open-ended questions. 
For each intervention (cheat sheets or Pattern Explainer), they were asked to describe how they used that intervention; how the intervention influenced the way they explored the network; how the intervention influenced the things they looked at in the network; and what they liked and disliked about using the intervention. 
We also asked them to compare the two interventions by reflecting on whether they used the interventions differently to explore the network, which one they preferred and why, and which one helped them learn more about network analysis and visualization. 
Besides, we collected their feedback about icon design, textual explanations, and UI for iterative improvement. 
Following this procedure, each session lasted around 60 minutes. Participants were given \pounds10 as compensation.

\subsubsection{Findings} 
Based on feedback from thinking aloud and interviews, we organize insights around the following themes: (1) pattern learning; (2) intervention usage; (3) preference; (4) issues raised; and (5) suggestions.

\textbf{Pattern learning.}
We observed that participants would describe their findings using network patterns and look for more patterns after having interventions, either cheat sheets or Pattern Explainer. 
In the first text-only condition, participants' descriptions were relatively low-level, echoing findings from the scoping study, e.g., explaining a single link, node size, and link color.
With interventions, they would search for high-level patterns (e.g., bridges, fans, and cliques), and use terms (e.g., \textit{``This is a hub''}).
P9 said, \textit{``I learn what they are actually called''}. 

Furthermore, the interactive Pattern Explainer was praised for filling the gap between abstract concepts and practical examples of patterns. 
For instance, P5 criticized that cheat sheets only showed graphic similarity, leaving semantics out, and thus, \textit{``if the patterns in the visualizations were quite different from those I saw in the cheat sheet, I would never relate them''} (P5). 
P5 felt that Pattern Explainer could help find more patterns. 
P6 and P12 mentioned that the real examples in Pattern Explainer helped them distinguish nuances between patterns. 
For example, without Pattern Explainer, participants found it hard to differentiate clusters and cliques in time-arcs, as well as bridges and hub nodes in matrices. After more learning, they mastered this by comparing specific examples and seeing more variants in the visualization.
As P12 commented, \textit{``the system puts the theoretic concepts into practice''}, aiding in interpreting patterns and exploring networks.

\textbf{Different intervention usage.}
We observed participants adopted different workflows using Pattern Explainer and cheat sheets. 
For Pattern Explainer, they commonly started with selecting an area, checked the information in the pop-up, and explored the examples. 
They described the process \textit{``intuitive''} (P3, P7, P9, P11, and P12), \textit{``integrated''} (P1, P2), and \textit{``faster''} (P1). 
P6 drew an analogy of Pattern Explainer and translator, saying \textit{``it provides in-time and in-situ explanations''}.
With straightforward interaction, people made diverse selections, e.g., visually salient parts, visual clutter, and white space. 
P9 explained \textit{``It took me to a place that I would never consider looking in the network...It guides me in that sense''}.
Besides, P2, P5, P7, and P8 valued that Pattern Explainer could validate their interpretation and build confidence by providing the correct answers.

For the cheat sheets, they usually read the content in the cheat sheet, remembered the visual explanations, looked for similar graphics in the visualization, and matched and validated by themselves.
P12 complained about the process of reading cheat sheets, \textit{``a bunch of information dumped to me at once''}, which needed them to figure out which was useful and existed by themselves, but they were afraid that they might make mistakes. 
P1 agreed: \textit{``Cheat sheets could not directly apply to the visualization. I got more confused about whether my understanding of this cheat sheet was right. Should I double-check the description? It makes me quite stressed''}.
P8 complemented, \textit{``Some patterns like parallel links rarely happen in my domain''}.

\textbf{Intervention preference.} 
Ten participants preferred Pattern Explainer to learn unfamiliar visualizations, as it gave in-context examples, recommended variations for exploration, helped interpret visually cluttered areas, and (in)validated assumptions. 
The two other participants favored cheat sheets.  
P8 argued, \textit{``The system lacks a clear progressive and organized learning path.''} 
The reason might be that they first used Pattern Explainer to interpret time arcs and the first explanation that popped out was about the fan pattern, which was relatively complex for one who had little knowledge about networks. 
While P3 preferred cheat sheets (\textit{``I could have my own learning pacing''}), they acknowledged that Pattern Explainer helped them learn more about network analysis and visualizations by seeing and exploring similar instances.

\textbf{Issues raised.}
One concern raised by the participants was potential over-reliance. 
While P5 said, \textit{``I [trust Pattern Explainer] very much, and it is very reliable''} and 
P1, P6, P7, P8, and P11 all mentioned they wanted to have Pattern Explainer all the time, these participants were also afraid that they would not be able to recall information without Pattern Explainer.  
Another issue is the inconclusive ``hit-and-miss'', where participants expected to have explanations but Pattern Explainer detected nothing. 
For example, P10 selected a portion that looked like a tree branch in the node-link diagram (\autoref{fig:visualizations}a), whereas the system just informed ``no patterns''. 
Other selections participants made but the system failed to provide expected explanations included white space (e.g., gaps between disconnected groups in node-link diagrams (\autoref{fig:visualizations}a) and plausible shapes (e.g., off-diagonal cells in matrices \autoref{fig:visualizations}b).
In that case, it cannot fully address participants' confusion.

\textbf{Improvements and suggestions.}
We iterated on the icon design and text based on participants' feedback. 
Major changes included: using contours to highlight hub and bridge nodes in matrices, only coloring a self-link in matrices, and differentiating connected groups of bridge nodes in time arcs.
We also received good suggestions from participants, like adding an animation/GIF~\cite{shu2020makes} to show how the examples in visualizations can be morphed into typical visual explanations (P6). 
\subsection{Quantitative Follow-up Study}
\label{sec:study3}

To complement our qualitative results and test assumptions in more detail,
we conducted a controlled user study to measure how many patterns people can accurately identify using either cheat sheets or interactive pattern explanations (between-subjects design). 
We focused on adjacency matrices since people are relatively familiar with node-link diagrams, and time-arcs were found too challenging previously. 

\subsubsection{Study Set-up}

\textbf{Participants}---We recruited participants from local universities with diverse majors. 
They were pre-screened according to their familiarity with adjacency matrices and node-link diagrams on a 5-point Likert scale (never seen (1) to expert (5)). 
20 participants were chosen having stated least familiarity: 18 participants had never seen adjacency matrices or did not know how to read them, and 2 participants had known but not used them before.

\textbf{Procedure}---We ran a between-subjects study, where 10 participants used Pattern Explainer with both bottom-up and top-down explanations and the other 10 used cheat sheets. 
After a brief introduction, each participant took part in a 15-min training session during which they used the assigned learning intervention to learn matrices while thinking aloud. 
After that, we asked participants to rate their confidence in the understanding of this visualization and its patterns on a 5-point Likert scale (very unconfident (1) to very confident (5)). 
We then took away the learning intervention and asked participants to identify and annotate as many patterns as possible in three other matrices on Figma, for 2 minutes each. After finishing, we asked for their confidence again. The session was audio- and screen-recorded and lasted around 30 minutes. Participants received \pounds6 for their participation.

\textbf{Data}---We used \textit{ `Les Mis\'erables' Co-occurrence} dataset for the training matrix, including 6 types of patterns: strong link, hub, bridge, fan, clique, and cluster. To keep the study controlled, cheat sheets only had these 6 patterns.
For the testing matrices, we chose a subset of the \textit{Stack Overflow Tag Network} dataset featuring 18 instances of the same six patterns. We manually generated two networks with 13 and 14 instances of six pattern types, and intentionally configured them with a certain complexity, having 53 and 54 nodes respectively.  
All networks had several instances of different patterns, as we aimed to simulate a natural exploration scenario where patterns were mixed with noise. 

\textbf{Measures}---
Two measures are chosen. First, two of the authors coded the videos based on the number of patterns participants identified correctly and in total. Ambiguities and mismatches were resolved through discussions.
For example, we only counted the strong link pattern once per matrix, as strong links could appear a lot and the number of strong links should not be dominant across all patterns. 
Second, we measured the pre- and post-confidence, instead of comparing the numbers of identified patterns before and after training. The reasons are that participants would receive informal training if they were asked upfront, bringing bias to the results. Moreover, subjective confidence can be a good measure since a person's confidence has a direct impact on their analysis, decision-making, and learning~\cite{boy2014principled}.

\subsubsection{Findings} 

\begin{figure}[tb]
    \centering
    \includegraphics[width=.95\columnwidth]{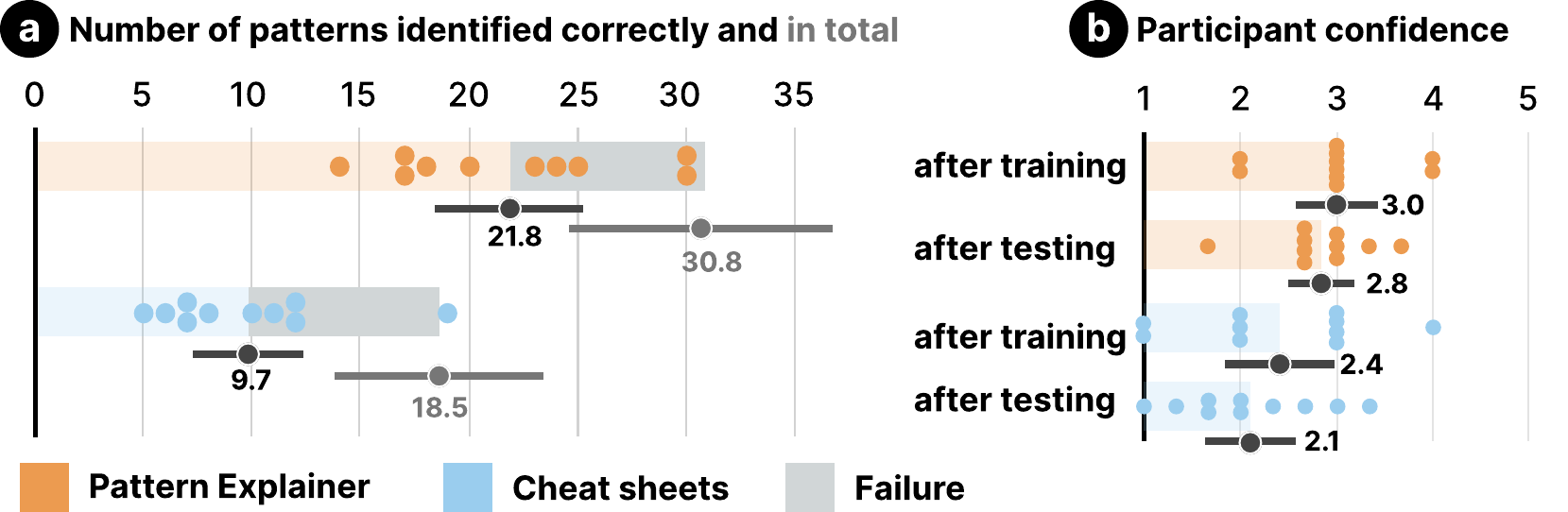}
    \caption{Quantitative study results. (a) shows the number of patterns correctly identified with the distribution as well as the number of patterns reported in total. (b) shows confidence after training and after testing. Error bars indicate 95\% confidence intervals.
    }
    \label{fig:study3}
\end{figure}

\autoref{fig:study3}a shows that participants trained in interactive Pattern Explainer correctly found an average of 21.8 patterns across all three matrices, compared to 9.7 patterns in average in the cheat sheet group. 
We found a large effect size of 2.48 (Cohen's d) and very small confidence intervals (95\% CIs, [7.2, 12.2] vs. [18.3, 25.2]), which translates into over 100\% more patterns detected with Pattern Explainer vs. cheat sheets (12.1 more patterns). Acknowledging the relatively small sample sizes, a Wilcoxon rank-sum test showed a significance of p=0.0006. Given the large effect size, we think those results
validate our hypothesis that interactive explanations increase the number of patterns found.

For those patterns reported in each group, we found a slightly higher accuracy means, namely 57\% for cheat sheets and 72\% for Pattern Explainer. However, the Pattern Explainer group reported 30.8 patterns on average, compared to 18.5 for the cheat sheets group. Thus, while similarly accurate, the cheat sheet group missed 40\% of the patterns the Pattern Explainer group reported. This gives a huge number of missed patterns (false negatives), in addition to the much smaller number of true positives for cheat sheets (\autoref{fig:study3}a).

In \autoref{fig:study3}b, participants after training on Pattern Explainer were more slightly confident (mean=3 on a scale of 1-5), compared to those trained on cheat sheets (mean=2.4). After testing, confidence dropped for either group, but slightly more so for the cheat sheet group (3$\rightarrow$2.8 vs. 2.4$\rightarrow$2.1
For example, two participants who rated themselves as confident after training in cheat sheets could only correctly identify a total of 7 and 8 patterns, respectively, in the testing visualizations. The follow-up discussions revealed this was because they misunderstood some patterns (e.g., bridge and clique). As cheat sheets cannot actively and exactly intervene with their perception, i.e., in the way Pattern Explainer does, participants could not correct their assumptions.

\autoref{tab:study3} shows the average number of pattern instances correctly identified by participants. Identifying cliques was easy, but none of the participants identified the clique shown as a split block. 
Participants cited the limited training time (15 mins) and the amount of patterns in the testing matrices, but were generally fine with the 2-min annotation time limit per matrix. 
Listening to participants' explanations as they annotated the visualizations, we found they tended to search for obvious visual patterns (e.g., cross, block) in the testing matrices and match them to network patterns. 
Although most participants did not have expertise in node-link diagrams, no one reported difficulties in understanding the network pattern using node and link representation in the black icons in the explainer pop-up. One participant commented \textit{``I understood bridges through the icon [in Network Pattern Explainer]''}.

\begin{table}[tb]
\resizebox{\columnwidth}{!}{%
\begin{tabular}{rlllllll}
Condition         & Sum & \#SL & \#Hub & \#Bridge & \#Fan & \#Clique & \#Cluster \\ \hline
Pattern Explainer & \textbf{21.8}         & 1.9                  & 2.4          & 2.7             & 1.1          & 8.0               & 5.7              \\ 
Cheat sheets      & \textbf{9.7 }         & 1.3                  & 0.9          & 0.7             & 0.7          & 3.6             & 2.5              \\ \hline
\end{tabular}
}
\caption{The average number of patterns participants correctly identified in three testing matrices in the quantitative study. SL: Strong links.}
\label{tab:study3}
\vspace{-0.3cm}
\end{table}
\section{Discussion}
We discuss the implications for interactive pattern explanation, technical solutions, and the wider role of visual patterns in visualizations.

\subsection{Interactive Pattern Explanation}
We proposed interactive pattern explanations to bridge the gap in visualization learning by supporting on-the-fly explanations of visual patterns.
The idea is demonstrated through a proof-of-concept interface, backed by existing algorithms and heuristics for pattern detection. 
Potentially, we could use any detection methods and heuristics, and extend the approach to other (network) visualizations.

Our main research question was to explore if interactive pattern explanations can help understand visualizations better, compared to pure text or cheat sheets.
Our studies confirmed this assumption, showing that \textbf{Pattern Explainer increases the number of patterns people identify correctly} compared to cheat sheets, highlighting that \textbf{non-experts appreciated the in-context and on-the-fly explanations} of abstract visual patterns and the fact that \textbf{abstract pattern ideas would have a terminology and could be applied to the actual data} they were tasked to understand. Thus, we recommend interactive pattern explainers as another tool for visualization onboarding~\cite{stoiber2022visualization} that has potentials to foster network (visualization) literacy~\cite{sayama2016essential,zoss2018network}.

Cheat sheets, in contrast, were perceived as a more comprehensive list of patterns to look up and gain an overview over the dictionary of patterns, their meaning, and their complexity. Still, many of our participants reported being overwhelmed by the information shown on cheat sheets and voiced difficulties mapping visual patterns they perceived in the visualization to the abstract concepts on the cheat sheets. So, we believe both approaches---\textbf{interactive pattern explanation and cheat sheets can be complementary techniques}, and we should explore ways to combine both in interfaces. 
For example, Pattern Explainer's top-down mode could be combined with cheat sheets to provide an overview of patterns and information in a visualization, or interactive cheat sheets that support in-situ explanations. 
Future studies can further compare the combined technique with the standalone ones.
Likewise, we could investigate ways to properly mine and consequently combine and analyze visual and network patterns, inspired by existing tools for motif analysis~\cite{schreiber2005mavisto,lekschas2017hipiler} and visual graph query interface~\cite{bhowmick2020aurora, yuan2022playpen}.

Composing explanations for patterns should also consider users' background knowledge and diverse domains.
One participant (P8) voiced confusion about the richness of the patterns in both Pattern Explainer and cheat sheets and wished for \textbf{more guided and progressive tutorials}. 
Meanwhile, different domains (history, social science, neuroscience, etc.) prioritize different patterns, have their own specific patterns, and most importantly attach specific meanings to individual patterns. While our Pattern Explainer aimed at generic network patterns, it is easy to include domain-specific explanations and terminology. However, work and explanations might be needed to fully explain these domain concepts but a challenge will remain to suggest explanations that match the very specific data set and its context.

Future work could also investigate the feasibility of \textbf{personalized tutorials} that explain visual patterns in a set of prepared examples, gradually building up from simple to more complex patterns, including more domain-specific terminology and examples in the explanations, e.g., about social networks or biomedical concepts. 
Future work should also take users' prior knowledge of different visualizations into consideration, e.g., asking users if they need more tutorials on node-link diagrams the first time when the explainer pops-up, or guiding users to learn from familiar visualizations to unfamiliar ones~\cite{ruchikachorn2015learning}.

\subsection{Technical Extensions and Generalizations}

Still, our repository of patterns (\autoref{fig:patterns}) is limited compared to the sheer number of structure types networks contain (\hyperref[sec:Ch3]{Ch3}), potentially including nodes, links, subgraphs, time, geolocation, and attributes. 
Specific domains are usually interested in specific patterns, e.g., feed-forward loops in biology~\cite{graphletBiology} or triangles in sociology. 
A comprehensive collection and classification of the most common artifacts and ambiguities is beyond the scope of this paper, but would be a helpful extension. Likewise, we could look into richer network motifs and information taking into account temporal or spatial information associated with network data as well as as sets and hypergraphs. This would imply studying other visualizations for networks~\cite{mcgeeStateArtMultilayer2019, beck2014state, zhao2016egocentric} and the visual patterns they produce. We hope that our work will help understand how to explain higher-level network patterns, e.g., through composition of lower-level patterns, their distribution, or other features explored in our work. While we focus on the direct translation from visual to network patterns, there are lots of avenues for future work.

In particular, there is a rich set of \textbf{visual and network patterns in temporal networks} where connectivity among nodes unfolds over time. The temporal characteristics further challenge defining and detecting both visual and network patterns. 
For example, we excluded most patterns from our initial repository, since the resulting visual patterns were not perceivable as visual patterns (Obscurings, \hyperref[sec:Ch4]{Ch4}) in \autoref{fig:patterns}c. 

That said, we strongly believe that interactive pattern explanation has \textbf{potential beyond network visualizations}. Such investigations need to address similar challenges as the ones we discussed in \autoref{sec:challenges}, namely to identify visual patterns in those visualizations, identify the corresponding data patterns, explore the mappings between visual and data patterns, and find ways to automatically translate between them. We have seen patterns explicitly discussed only for geographic data and map visualizations~\cite{andrienko2022seeking}. We believe Pattern Explainer could play a crucial role in investigating artifacts and ambiguities in more complex and information-rich visualizations. 
Literature explicitly mentioning visual patterns, e.g., for temporal data~\cite{bach2015time}, spatio-temporal data~\cite{bach2017descriptive}, or treemaps and parallel-coordinates plots~\cite{wang2020cheat} can be a good start.

\subsection{Visual Patterns: A Critique}

The main challenge for any research into explaining visual patterns and learning visualization techniques is to \textbf{devise valid ways to evaluate the success of any intervention}~\cite{bach2023challenges}. 
One of the main characteristics of visual patterns is their fuzziness and variation, both structural and visual---up to the extent where visual patterns would be entirely indistinguishable or meaningless (\hyperref[sec:Ch4]{Ch4}). 
Pattern Explainer was motivated exactly to help resolve ambiguity and fuzziness. As a result, we opted for more open-ended ways to understand how people used the Pattern Explainer and if they found the approach and explanations useful to understand the visualization. Built on the results in this paper, future studies could observe people using interventions like cheat sheets or Pattern Explainer over a longer period of time, ideally while engaging in their own analysis~\cite{alkadi2022understanding}. Such a study could help provide empirically-grounded insight into which visual pattern people find useful (bottom-up), which network pattern they are interested in (top-down), and how personal pattern repertoires form and change over time. 

A risk in over-emphasizing the use and training for (visual) patterns might be the \textbf{danger of developing a narrow mind} and \textbf{over-reliance on the system}. There might be good reasons why the literature on visualization has not been discussing patterns in a more formal sense and why it has been restrained from defining pattern repositories. The open and fuzzy nature of patterns is what might make them useful---as a thinking tool and conceptual aid to approach teaching visualization techniques and thinking with visualization---rather than a formal dogmatic way of parsing visualizations. Visualizations are hard to describe by composition and dissection: they need to be open in the data patterns they want to show instead of prescribing what to show; they need to give \textit{some} perspectives on the data, show messiness where there is in the data and invite for human interpretation and inquiry---new visual patterns might emerge from the rules of the visual design and people will realize those patterns exist. 
In the best case, humans, after learning these initial patterns with Pattern Explainer, can question and refine their mental model for visualization, thinking and reasoning. 
Future visualization learning systems can be further strengthened with functionalities that engage users in critical thinking and active learning, such as asking follow-up questions~\cite{tanahashi2016study}, sketching and comparing visual patterns, thereby enhancing the internalization of the pattern concepts and mitigating the influences of over-reliance on learning.

In conclusion, we \textbf{could ponder a more comprehensive `critical theory of patterns'} in visualization. Not a theory that meticulously wants to cartograph any existing pattern or create a dogmatic framework. Rather, a theory that facilitates to describe, analyze, evaluate, and communicate visual patterns in visualization and provides the necessary terminology and methodological tools. A theory that describes visual patterns as essential building blocks of (complex) visualizations and offers keys to their interpretation. We could use such a concept of patterns to scrutinize a visualization (design) and algorithm and assess both visual artifacts and hallucinators; we could think about ways to make specific patterns more salient in a given design by tweaking its design; we could suggest visualization designs that are best suited to show specific patterns; and we could think about guiding novices in understanding visualizations and interactive visual analysis. 
Such a theory could become a structured way to engage visualization design and visualization education in a \textbf{critical and creative way} and to show the potentials and pitfalls of data visualization more generally.

\acknowledgments{
The authors thank the participants for their help and the anonymous reviewers for their valuable comments. This work was supported by a grant from EPSRC (Project EP/V010662/1).%
}

\newpage
\bibliographystyle{abbrv-doi-hyperref}

\bibliography{main}

\end{document}